\documentclass[showpacs,amsmath,amssymb,prd,onecolumn,nobibnotes,12pt]{revtex4-2}

\usepackage{ulem}
\usepackage{latexsym}
\usepackage{graphicx}
\usepackage{slashed}
\usepackage{pifont}
\usepackage{color}

\newcommand{\req}[1]{Eq.\,(\ref{#1})}
\newcommand{\beqn}{\begin{equation}}
\newcommand{\eeqn}{\end{equation}}

\begin{document}
\title{Kinematically boosted pairs from the nonlinear Breit-Wheeler process in small-angle laser collisions}
\author{Pisin Chen}
\affiliation{Leung Center for Cosmology and Particle Astrophysics\\
National Taiwan University, Taipei, 10617 Taiwan}
\affiliation{Department of Physics and Graduate Institute of Astrophysics\\ 
National Taiwan University, Taipei, 10617 Taiwan}
\affiliation{Kavli Institute for Particle Astrophysics and Cosmology\\
SLAC National Accelerator Laboratory, Menlo Park, CA 94025, USA}
\author{Lance Labun}
\email{Correspondence: labun@utexas.edu}
\affiliation{Leung Center for Cosmology and Particle Astrophysics\\
National Taiwan University, Taipei, 10617 Taiwan}
\affiliation{Department of Physics, The University of Texas, Austin, 78712, USA}
\affiliation{Tau Systems, Inc., Austin, Texas, 78701, USA}
\date{April 20, 2023} 

\begin{abstract}
We discuss a scheme of nonperturbative pair production by high energy photons ($\omega\gtrsim m$) in a strong external field is achievable at the next high intensity laser experiments.  The pair momentum is boosted and for $\omega\gtrsim 1.2m$ the pair yield is increased when the external field is formed by two laser pulses converging at a small angle.  These characteristics are nonperturbative in origin and related to the presence of magnetic field in addition to electric field.  By enhancing the signal over perturbative backgrounds, these features allow the employment of above-threshold photons $\omega>2m$, which further increases the pair yield. We note the close relation of this photon-pair conversion mechanism to spontaneous pair creation, recommending it as an accessible stepping stone experiment using state-of-the-art or soon-to-be laser technology.
\end{abstract}

\maketitle
\section{Introduction}
In quantum electrodynamics (QED), spontaneous pair production is the emission of electron-positron pairs by a low frequency $\hbar\omega\ll m_ec^2$ electromagnetic field.  It has been recognized as the hallmark of nonperturbative QED since Sauter \cite{sauter1931verhalten}, Heisenberg, Euler \cite{heisenberg1936folgerungen} and Schwinger \cite{schwinger1951gauge} achieved some of the first calculations~\cite{greiner2012quantum,ritus1985quantum,chen1999boiling,dunne2008new,di2012extremely,salgado2021towards,fedotov2022advances}.  Observation of this QED process would shed light on related nonperturbative quantum field theory (QFT) processes such as the flux-tube model of particle creation in hadron collisions \cite{andersson1998lund} and Seiberg-Witten brane-anti-brane pair creation in string theory \cite{seiberg1999d1, Ong:2013bia}. 
The nonperturbative production probability is exponentially suppressed  by the field magnitude $|\vec E_c|=m_e^2/e\simeq 1.32\times 10^{18}$\,V/m ($\hbar=c=1$ units).  For a single-mode oscillating field, the exponential is continuously related to the nonlinear, $N$-photon process~\cite{brezin1970pair}, which goes as a power law~\cite{brezin1970pair,nikishov1964quantum,narozhnyi1964quantum, nikishov1965nonlinear,nikishov1967pair}.
The SLAC E-144 experiment achieved the multiphoton process with $N\simeq 5$~\cite{bula1996observation, bamber1999studies}, and observation of the nonpertubative $N\gg 1$ process is now a realistic goal thanks to progress in high intensity laser technology.

The ELI~\cite{eliwhite} and ZEUS \cite{nees2020zeus} facilities are expected to achieve laser intensities corresponding to $|\vec E|\sim (10^{-4}-10^{-3})|\vec E_c|$, nearing but still a few orders of magnitude from the ``critical'' field at which spontaneous production is rapid~\cite{chen1999boiling, labun2009vacuum}.  For this reason, a variety of ways to increase the production rate have been studied \cite{dunne2008new,di2012extremely, salgado2021towards, fedotov2022advances}.  One approach is to introduce photons, which can convert into pairs when propagating in a high intensity external field \cite{ritus1985quantum, dittrich2000probing, dunne2009catalysis, krajewska2012breit, nousch2012pair, Titov:2012rd, karbstein2013photon, titov2013breit, wu2014nonlinear, meuren2015polarization, 
di2016nonlinear, otto2016pair, blackburn2018nonlinear, mackenroth2018nonlinear, king2020uniform, seipt2020spin, wan2020high, mercuri2021impact, song2021spin, tang2021pulse, gao2022optimal, golub2022nonlinear, tang2022fully, macleod2023all}.  Since the wavenumber of the external field is much smaller than the electron mass, pair conversion of a single photon requires absorption of $N\gg 1$ quanta from the external field, meaning it is nonperturbative even when the photon $\omega>2m$. As discussed below, non\-per\-turbative pair conversion shares analytic structure with spontaneous pair production, but differs enough to be achievable at near-future laser facilities, offering an experimental stepping stone of independent theoretical interest.

However, there are limitations to the simple setup of photons in a pure electric field created by counter-propagating laser pulses \cite{dunne2009catalysis}.  The yield of pairs is proportional to the number of high frequency photons, and a large number of photons ($10^{10}$) must be injected to compensate the exponential suppression when the laser field strength is significantly below $|\vec E_c|$.  At the photon number density corresponding to packing those photons in the few-micron scale focal volume of an ultra-high intensity laser, perturbative pair production ($\gamma\gamma\to e\bar e$) becomes probable whenever the invariant mass of two photons passes the kinematic threshold, $s=k_1\!\cdot\!k_2<4m_e^2$ where $k_{1,2}$ are the 4-momenta of the two photons.  The possibility of perturbative production suggests we should also consider phenomenological features that distinguish nonperturbatively produced pairs from perturbatively produced pairs.

In this work, we show how to simultaneously enhance nonperturbative pair conversion and give the produced pairs a characteristic large momentum (rapidity) that is determined by the geometry of the high intensity lasers.  Large pair rapidity is achieved by boosting the center of momentum frame of the process: nonperturbatively produced pairs inherit momentum from the external field, as seen by considering it diagrammatically as absorption of $N\gg 1$ soft photons~\cite{chiu1979diagrammatic}.  For spontaneous production, the center of momentum frame of produced pairs coincides with the rest frame of the external field~\cite{labun2011spectra}.  For pair conversion, the high energy photon also contributes momentum.  To maximize this kinematic boost, we set two high intensity laser pulses to converge at a small angle $\phi$.  The superposed laser fields create a total field that is both off the photon shell (necessary for nonperturbative pair creation) and at a high momentum relative to the lab, controlled by $\phi$.  
A novel aspect when considering nonperturbative pair conversion is that for photon frequency near threshold $\omega \gtrsim m_e$, small $\phi$ increases the number as well as the momentum of the produced pairs.  

This scheme has several advantageous features. First, the relationship between pair momentum and the control parameter $\phi$ offers a signature to establish the nonperturbative origin of the pairs.  Second, creating pairs at high momentum may help control the kinematics of secondary production, that is cascades~\cite{fedotov2010limitations,elkina2011qed}.  If so, we improve the chances to identify momentum signatures characteristic of nonpertubatively produced pairs~\cite{hebenstreit2009momentum, orthaber2011momentum, fey2012momentum}. Third, the kinematic boost combines with other discovered phenomena enhancing the yield, such as short-pulse effects \cite{krajewska2012breit,Titov:2012rd}.

\section{Nonperturbative pair conversion}
To help explain the mechanism enhancing pair conversion, we first recall features of the spontaneous production mechanism, which is applicable to fields with frequency $\omega\ll m$.  The expected number of pairs per unit volume per unit time is given by the first term of the Schwinger series~\cite{Nikishov:1970br,Cohen:2008wz}
\beqn\label{SchwingerN}
\frac{N_{ee}}{VT}=\frac{a^2}{4\pi^3}\frac{b\pi}{a}\coth\!\left(\frac{b\pi}{a}\right)e^{-\pi m_e^2/a}.
\vspace*{-1mm}
\eeqn
The rate depends on the invariants, 
\vspace*{-1mm}
\begin{align}\label{abinv}
a^2&=e^2(\sqrt{\mathcal{S}^2\!+\!\mathcal{P}^2}-\mathcal{S}), \cr
-b^2&=-e^2(\sqrt{\mathcal{S}^2\!+\!\mathcal{P}^2}+\mathcal{S}),
\end{align}
which are the squared eigenvalues of the field tensor $eF^{\mu\nu}$ written in terms of the scalar and pseudoscalar invariants
$2\mathcal{S}=(\vec B^2\!-\!\vec E^2)$ and $\mathcal{P}=-\vec B\cdot\vec E$.  
Here $a,b$ are the electric and magnetic field strengths in the field rest frame, which explains why $a$ appears in the exponent.  The presence of magnetic field aligned with the electric field enhances the rate, seeing as $x\coth x\geq 1$ with equality only in the $x\to 0$ limit. However $a,b$ are constrained if both electric and magnetic fields are supplied by laser pulses. 

In the presence of magnetic field, the momentum density of the electromagnetic field is in general nonvanishing.  Consequently, the Lorentz frame in which spontaneously created pairs appear is moving with respect to the laboratory.  This feature enables boosting the spontaneously produced pairs, but also implies  a trade-off between the energy and the yield of the produced pairs~\cite{labun2011spectra}, which will be elaborated below. 

In contrast, magnetic field assisted pair conversion can enhance the overall discovery potential without sacrificing pair yield, because it involves a new invariant
\begin{align}\label{chiinv}
\chi^2\:& =|eF^{\nu}_{\mu}k^{\mu}|^2 
\\ &\to |e\vec E|^2(k_0^2-k_z^2)+|e\vec B|^2(k_x^2\cos^2\theta+k_y^2+k_z^2\sin^2\theta). \notag
\end{align}
In the second line we evaluate the invariant in a coordinate system with $\vec E$ aligned in the $z$-direction and $\vec B$ at an arbitrary angle $\theta$ in the $x\!-\!z$-plane.  $\chi$ is maximized when the photon travels in the $x$ or $y$ directions, and if in the $x$ direction, we should choose $\theta=0$, i.e., $\vec B\Vert\vec E$.  With this choice, $\chi$ reduces to the product of the field energy density and the photon frequency. 

Pair conversion is described by the imaginary part of the photon polarization tensor $\Pi^{\mu\nu}$ evaluated in an external field.  
Seeking the total pair yield, we average over photon polarizations, which just requires the trace of the polarization tensor.  This saves diagonalizing the polarization tensor in a general external field with both $\vec E,\vec B\neq 0$, but introduces an $\mathcal{O}(0.1)$ error in the yield considering that the source of photons may be partially polarized.  This also emphasizes that the coherence length of the process $\ell_f\sim m_e^{-1}$ \cite{Baier:2003hf} is much smaller than the spatial variation of the field.  For this reason, we can neglect the frequency-dependence of laser fields and use the general form of $\Pi^{\mu\nu}$ for quasi-constant ($\omega\ll m_e$) fields, given in~\cite{dittrich2000probing}.  We need only the two transverse components in the tensor decomposition, because the longitudinal components do not contribute to trace: the $0$ component is assumed to have no nontrivial solutions to the lightcone condition $k^2+\Pi_0=0$, hence no propagating modes, and the projection tensor associated to the third space-like component is zero under the trace.

We define the polarization-averaged inverse absorption length
\beqn\label{kappadefn}
\bar\kappa=-\frac{1}{\omega}\frac{1}{2}\sum_{\sigma}\Im\:\Pi_{\sigma}
\eeqn
where $\sigma=\Vert,\perp$ runs over transverse polarizations.  The expectation value of the absorption probability is then the exponent of $\kappa$ times the distance $L$ the photons propagate in the external field, and the number of pairs produced equals the number of photons absorbed, 
\beqn\label{Nee}
N_{ee}=(1-e^{-\bar\kappa L})N_\gamma. 
\eeqn

For the case $\vec B=0$, $\Im\Pi$ is evaluated to high accuracy using contour integration and the saddle-point approximation to resum the poles~\cite{dunne2009catalysis}.  Nonzero magnetic field does not change the analytic structure of the polarization tensor, leaving the poles in the same location but modifying their residues, as in the case of the effective action \cite{schwinger1951gauge}.  For this reason, we can adapt the same procedure, and continuity with the previous calculation is demonstrated by verifying that the limit $b\to 0$ reproduces at each intermediate step the results of Ref.~\cite{dunne2009catalysis}.  Much below the critical field, the contour integration is sharply peaked near the saddlepoint (as seen in other sfQED calculations \cite{ritus1985quantum}), which is why the approximation yields a high accuracy result here and in the previous work. The final result is
\begin{align}\label{ImPifinal}
\sum_{\sigma}\Im\Pi_\sigma=
&\frac{\alpha a}{2}\left(\left|s_*(\frac{bs_*/a}{\sin(bs_*/a)}-\frac{s_*}{\sinh s_*})\right|
\left|\frac{(b/a)\sin bs_*/a}{\cos^4(bs_*/2a)}+\frac{\sinh s_*}{\cosh^4(s_*/2)}\right|\right)^{-1/2}\\
&\times\frac{e^{-i\Phi s_*/a}(bs_*/a)}{\sinh s_*\sin (bs_*/a)}\big(N_*(b/a,1)-N_*(1,ib/a)\big) \notag 
\end{align}
with
\begin{align}
\notag \Phi=m^2&-\frac{v_\perp^2}{2}\frac{\cos\nu bs-\cos bs}{bs \sin bs}+\frac{v_\Vert^2}{2}\frac{\cosh\nu as-\cos as}{as \sinh as}, \\
 &N_*(x,y)=2\cos(xs_*)\frac{\cosh(ys_*)-1}{\sinh^2 (ys_*)}\,. 
\end{align}
The scalars $v_\perp^2,v_\Vert^2$ are derived from the invariant decomposition of the photon momentum vector \cite{dittrich2000probing}, 
\begin{align}
v_\perp^2&=\frac{b^2k^2+\chi^2}{a^2+b^2}\\
v_\Vert^2&=\frac{\chi^2-a^2k^2}{a^2+b^2}.
\end{align}
Even in an external field, the deviation from the vacuum lightcone condition $k^2=k^\mu k_\mu=0$ is less than $10^{-3}$ \cite{dittrich2000probing}.  For comparison, the invariant $\chi^2$ scales with the squared magnitude $|\vec k|^2$ of the photon wave number, except in the special case that all three vectors are parallel $\vec k\Vert\vec E\Vert\vec B$ where it vanishes.  Therefore, the approximation that $k^2=0$ in the external field, collapses the two scalars into one $\tilde v^2$,
\beqn
v_\perp^2=v_\Vert^2
=\frac{\chi^2}{a^2+b^2}\equiv \tilde v^2,
\eeqn 
and introduces an error smaller than $10^{-3}$ in its evaluation.
All expressions in \req{ImPifinal} are evaluated at the saddlepoint $s_*$, which we solve for numerically as the solution to the transcendental equation
\beqn\label{sp}
\frac{1}{1+\cosh s_*}+\frac{2m^2}{\tilde v^2}=\frac{1}{1+\cos bs_*/a}
\eeqn

In the exponential factor $\exp(i\Phi s/a)$, the leading contribution in the low frequency limit $\omega\to 0$ is the first $m^2$ term, which produces $e^{-\pi m^2/a}$ dependence like \req{SchwingerN}.  For larger values of the invariant $\chi$, however, the second and third factors $\propto \chi^2/a^2$ can compensate small $a/m_e^2$.  

Using \req{ImPifinal}, we have evaluated the pair yield with different external field geometries.  Similar to the spontaneous production case \req{SchwingerN}, nonzero $b$ enhances the total yield.  Although $b$ is present in the exponent $\Phi$ and the enhancement grows faster than linear for large $b/a$, the pair yield is again maximized by optimizing the field for the $a$ invariant.  Moreover, the limit $b\to 0$ is smooth, meaning small $b\neq 0$ is a small positive correction to pair yields.

\section{Boosted pair production}
To determine the momentum of the produced pairs, we first go to the rest frame of the high intensity field, which has been treated as classical in the preceding calculation of pair production.  The field rest frame is the frame in which its 3-momentum vanishes, with the electromagnetic 4-momentum defined covariantly from the energy-momentum tensor 
\beqn\label{Pmu}
P^{\mu}=T^{\mu\nu}u_\nu 
\underset{u_\nu=(1,\vec 0)}{\longrightarrow}
\begin{cases}
T^{00}=\frac{1}{2}(\vec E^2+\vec B^2) \\
T^{0i}=\vec E\times \vec B
\end{cases}
\eeqn
where $P^{\mu}$ is the  momentum density and $u_\nu$ is a 4-vector defining the observer~\cite{landau1971classical}.  Taking $u_\nu=(1,\vec 0)$ means the observer is at rest in the Lorentz frame being considered, and then this definition produces the usual Poynting vector.  The field momentum and rest frame are quasi-local quantities, which should be considered as integrated over a mesoscopic volume chosen smaller than the length scale over which the field varies, but larger than the length scale associated with pair formation.  Since the laser and pair production scales are widely separated, with the laser wavelength $\lambda_{\rm laser}\sim(1\,\mathrm{eV})^{-1}$ much greater than the pair formation length $\ell_f\sim m_e^{-1}$, we can clearly choose a mesoscopic length scale $\ell$ satisfying $\lambda_{\rm laser}\gg \ell \gg m_e^{-1}$ which defines the quasi-local volume.
In strong fields $|\vec E|\sim m_e^2/e$, QED significantly modifies the Maxwell energy-momentum tensor~\cite{labun2010dark}; however, aiming at the next generation of experiments attaining fields $|\vec E|< 0.1|\vec E_c|$, we can omit these corrections.

The magnitude $P^{\mu}P_{\mu}$ is invariant
\beqn\label{massdensity}
\sqrt{P^{\mu}P_{\mu}}\equiv\mu=\frac{1}{2}(a^2+b^2) 
\eeqn
showing the energy density in the rest frame (the mass density) depends only on invariants \req{abinv}.  
The transformation to the field rest frame is obtained from the condition that $\vec E'\times\vec B'=0$, prime denoting quantities in the rest frame.  Plugging in the Lorentz transformation for $\vec E',\vec B'$ in terms of $\vec E,\vec B$ and making the Ansatz,
$ \vec\beta=C(\vec E\times\vec B) $,
we find a quadratic equation for $C$ with two solutions $C^{-1}=T^{00}\pm\mu$ where $\mu$ is the mass density defined in \req{massdensity}.  The requirement $\beta^2<1$ means only the (+) solution is physical, and the boost velocity to the field rest frame is
\beqn\label{boostvel}
\vec\beta=C(\vec E\times\vec B), \quad C=(T^{00}+\mu)^{-1}
\eeqn
Since for light-like fields, the energy density equals the momentum density $T^{00}=|\vec E\times\vec B|$, we see that smaller $\mu$ means a larger boost, $\beta^2\to 1$.  

For a simple case to study the boost, we consider two converging laser pulses with equal intensity and frequency, and momentum vectors satisfing $\vec P_1\cdot\vec P_2\propto\cos\phi>0$.  Due to the exponential suppression, pair production is significant only in regions where the (total) field invariants are maximized.  We calculate the invariants of the total field as from converging plane waves: by comparison to a realistic pulse model~\cite{narozhny2000scattering,narozhny2004e+}, corrections are subleading in the small focusing parameter $\Delta=\lambda_{\rm laser}/2\pi R$, where $\lambda_{\rm laser}$ is the laser wavelength and $R$ the radius of the pulse waist.  %
In the overlap region of two converging plane waves, either the magnitude of the net electric field or net magnetic field is larger, depending on how closely aligned the plane formed by $\vec P_1,\vec P_2$ is with the plane formed by the two polarization vectors $\vec E_i/|\vec E_i|$.  We assume the polarizations of the laser pulses are chosen to maximize $\mathcal{S}$, so as to maximize pair yield, according to the discussion above.  In this case, 
$\mathcal{S}=2|\vec E|^2\sin^2\phi,~\mathcal{P}=0$.
These values are exact for two converging plane waves, and valid to leading order in $\Delta$ for (quasi-)circularly polarized laser pulses~\cite{narozhny2004e+}.  Since $a^2=|\mathcal{S}|$, the invariant $a$ in the exponent in \req{SchwingerN} is small in the interesting limit of small $\phi$, which suppresses spontaneous production. 

The invariants of the combined laser fields give $\mu=2|\vec E|^2|\sin^2\phi|$, so that, as expected, a smaller convergence angle (more light-like total field), means a larger boost to the rest frame and a larger rapidity for produced pairs.  Lastly, we need the momentum of the produced pairs as they appear in the rest frame of the field.  To achieve the highest boost, we inject the high energy photons copropagating with the net momentum of the high intensity field.  Considering the pair creation process in the field rest frame, we observe that the photon has frequency and momentum $\omega'=|\vec k'|\ll m_e$.  With $\vec k$ perpendicular to $\vec E$, the momentum of the tunneling state is transverse to the field providing the tunneling potential.  Therefore, in the relevant adiabatic limit, the pair materializes with zero longitudinal momentum, $p_\Vert=0$, the $\Vert$ direction defined by the $\vec E$ field vector.  The mean value of the transverse momentum is determined by momentum conservation as the momentum of the photon $\langle p_{\perp}\rangle=\vec k'$ (evaluated in the field rest frame).  

Using additivity of rapidities, we find the mean rapidity (Lorentz factor $\gamma_{ee}$) of the produced pairs
\beqn\label{gammaee}
y_{ee}=y_F+\sinh^{-1}\left(\!\sqrt\frac{1\!-\!|\vec\beta|}{1\!+\!|\vec\beta|}\frac{\omega}{m}\right), 
~~ \gamma_{ee}=\cosh y_{ee}
\eeqn
where the field rapidity is $y_F=\cosh^{-1}(1/\sqrt{1-\beta^2})$, $\vec\beta$ given in \req{boostvel}.  Here $\omega$ is the photon frequency in the lab frame, and the cofactor in the argument of arcsinh is the Doppler factor for the shift to the field rest frame.  For large boosts, i.e. small $\phi$, the second term is subleading, and $\gamma_{ee}\simeq\gamma_F\sim 1/4|\sin^2\phi|$.

\begin{figure}
\includegraphics[width=0.48\textwidth]{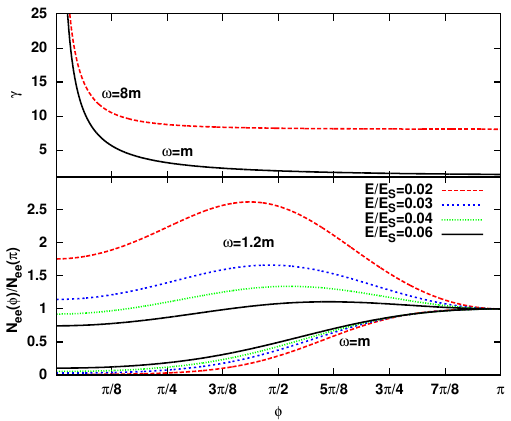}
\caption{ Upper panel: the Lorentz factor of the produced pairs \req{gammaee} as a function of convergence angle $\phi$.  Lower panel: the relative yield of pairs, normalized to the counter-propagating case $\phi=\pi$, for different seed photon frequencies, $\omega=m$ (lower family of curves) and a transitional  case $\omega=1.2m$ (upper family), above which pair yield is enhanced at smaller $\phi$ (see also Figure~\ref{fig:Nee2}). \label{fig:boost} }
\end{figure}

Figure~\ref{fig:boost} shows the pair Lorentz factor as a function of convergence angle of the two laser pulses.  For comparison, we plot the relative pair yield, normalized to the yield for head-on pulses $\phi=\pi$.  The angle dependence is sensitive to the seed photon energy.  At $\omega\lesssim m$, head-on pulses produce the highest yield, because  smaller $\phi$ reduces $a$.  For $\omega\gtrsim 1.2m$, the yield increases for small $\phi$, because decreasing $a$ increases $\tilde v^2/m^2$, which becomes dominant in determining the pair yield.  

For $\phi=0$, the two laser pulses are exactly co-propagating and can be described by a single light-like field distribution.  In this case the field tensor $F^{\mu\nu}$ can be written in terms of the antisymmetrized product of a tranverse (polarization) vector and a lightcone unit vector $n^\mu=(1,0,0,1)$.  The 4-momentum of a photon co-propagating with the laser field(s) is also proportional to $n^\mu$, $k^\mu=\omega n^\mu$. Consequently, the invariant controlling pair conversion, $\chi$ \req{chiinv}, vanishes exactly, as an example of Schwinger's theorem that no nonlinear vacuum phenomena occur in plane waves \cite{schwinger1951gauge}.  It is important to understand that the pair conversion vanishes only for $\phi=0$.  For any non-zero $\phi$, the invariant is non-zero, whether the classical field is modeled as a plane wave or a quasi-constant field.  Therefore, the limit $\phi\to 0$ is not uniform.  Relating to a distribution of measure zero ($\delta(\phi)$), this fact is not easily made manifest in the figure and is not of great physical importance considering that laser fields are focused generating nonvanishing field invariants and pair creation \cite{narozhny2004e+}.

The impact of high pair momentum on cascade development is seen considering the radiation length~\cite{fedotov2010limitations,elkina2011qed} 
\beqn
\xi_{ee}=\alpha^{-1}p_0\chi^{-2/3}\simeq 2\times 10^{-4}\gamma_{ee}^{2/3}~\mu\mathrm{m}
\eeqn
with $p_0$ the electron or positron energy and $\chi$ from \req{chiinv}.  For a relativistic electron traveling (initially) orthogonal to the electric field, $\chi\sim|\vec E|p_0$ and we obtain the scaling relation on the right.  To suppress cascades, we must have $\xi_{ee}>\lambda\sim 1\:\mu\mathrm{m}$ the laser pulse length scale.  This requires $\gamma_{ee}\gtrsim 10^7$, corresponding to $\phi\sim 10^{-3}$.  Even achieving $\xi_{ee}\sim 0.1\lambda$ should significantly reduce cascade development.  Achieving the high gamma factor facilitates search for the momentum signatures associated with spontaneously produced pairs~\cite{labun2011spectra, hebenstreit2009momentum, orthaber2011momentum, fey2012momentum}, as well as being interesting in its own right for the production high energy electron bunches.

\begin{figure}
\includegraphics[width=0.48\textwidth]{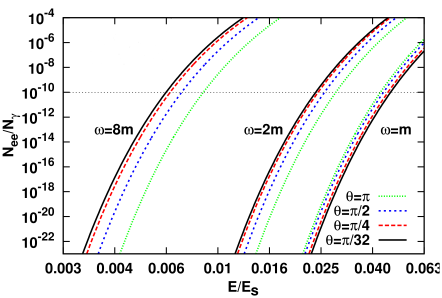}
\caption{ Pair yield, normalized to input photon number, as a function of field strength for different values of the photon frequency and convergence angle.  The length of the strong-field region is taken to be $L=0.75\,\mu$m.} \label{fig:Nee2} 
\end{figure}

For yield estimates, we consider example parameters based on ELI preliminary performance reports.  In estimating the propagation length $L$ in \req{Nee}, we note that for counter-propagating pulses, the relevant length is focal spot diameter, which could be as small as $\sim 2 \mu$m to maximize in the intensity.  On the other hand, for small angle collisions with 10 PW--250 J pulses, the relevant length is the pulse duration or twice the Rayleigh range, both of which lead to $L\sim 7.5\,\mu$m long.  On the other hand, below the critical field, the yield is much less than one pair per photon, so that $1-\exp(-\bar\kappa L)\simeq \bar\kappa L$ showing that the yield increase from the geometric increase in propagation length with decreasing convergence angle is at most a factor 5.  To be conservative and isolate the kinematic effect on the yield, we take a length about half the spot diameter, $L=0.75\,\mu\mathrm{m}$, and drop the order 1 $\phi$-dependent geometric factor.
In figure~\ref{fig:Nee2}, pair yields normalized to the number of high frequency photons $N_\gamma$ are shown as a function of laser field strength, for different values of seed photon frequency and laser convergence angle.  This shows in absolute scale the advantage in having near threshold seed photons: yield decreases with $\phi$ for $\omega\leq m$, but increases with $\phi$ for $\omega\gtrsim 1.2m$.  We can consider above threshold photons, such as $\omega=8m$, in conjuction with a small convergence angle, since we may be able to identify nonperturbatively produced pairs by their large initial energy.

\section{Conclusions}
To summarize, we have calculated nonperturbative photon-pair conversion in converging laser pulses.  This configuration takes advantage of the exponential enhancement due to the high frequency seed photon $\omega\gtrsim m$ at the same time as boosting the momentum of the produced pairs.  Since the pair energy is directly related to the experimental control parameter, the laser convergence angle $\phi$, the correlation $\gamma_{ee}\sim \phi^{-2}$ provides an identifying feature of nonperturbatively produced pairs.  
We have found that for photons with $\omega\gtrsim 1.2m$, decreasing the convergence angle significantly enhances the pair yield over the counter propagating case.  The yield enhancement is over 6 orders of magnitude at small convergence angles, with $\omega=8m$ and electric field strength 0.004$|\vec E_c|$, corresponding to a $\sim 215$ PW pulse at the demonstrated ELI pulse duration of $22$ fs \cite{radier202210} and assuming a realistic Strehl ratio of 0.75 for a 1 $\mu$m Gaussian spot radius.  Such a significant frequency-dependent enhancement would strengthen the signal in the context of a broad distribution of photon frequencies such as provided by a bremsstrahlung source \cite{blackburn2018nonlinear, eckey2022strong}. The $\phi$-dependent momentum boost provides a signature to identify the nonperturbatively produced pairs from possible backgrounds.  

On the other hand, the relative scalings of the pair-conversion and spontaneous processes must be studied quantitatively to determine an optimum experimental strategy~\cite{bulanov2011design}.  For instance, the pair conversion process depends on the length $L$ of the high intensity field, whereas the spontaneous process scales with volume $L^4$.  Moreover, some laser energy must be diverted to create high frequency photon bunch.   By enhancing the pair-conversion yield, our scheme strengthens the case for this avenue toward discovering nonperturbative pair production, especially since pushing a little farther to threshold $\omega=2m$ provides direct laboratory access to the pair-conversion process essential in high energy astrophysics.

{\it Acknowledgments:}  LL is grateful to thank C. Klier and J. Rafelski for discussion improving the definition of the field rest frame.  Pisin Chen is supported by Taiwan National Science Council under Project No. NSC 97-2112-M-002-026-MY3 and by US Department of Energy under Contract No. DE-AC03-76SF00515.

\bibliographystyle{apsrev4-2}
\bibliography{pairs}

\begin{thebibliography}{63}%
\makeatletter
\providecommand \@ifxundefined [1]{%
 \@ifx{#1\undefined}
}%
\providecommand \@ifnum [1]{%
 \ifnum #1\expandafter \@firstoftwo
 \else \expandafter \@secondoftwo
 \fi
}%
\providecommand \@ifx [1]{%
 \ifx #1\expandafter \@firstoftwo
 \else \expandafter \@secondoftwo
 \fi
}%
\providecommand \natexlab [1]{#1}%
\providecommand \enquote  [1]{``#1''}%
\providecommand \bibnamefont  [1]{#1}%
\providecommand \bibfnamefont [1]{#1}%
\providecommand \citenamefont [1]{#1}%
\providecommand \href@noop [0]{\@secondoftwo}%
\providecommand \href [0]{\begingroup \@sanitize@url \@href}%
\providecommand \@href[1]{\@@startlink{#1}\@@href}%
\providecommand \@@href[1]{\endgroup#1\@@endlink}%
\providecommand \@sanitize@url [0]{\catcode `\\12\catcode `\$12\catcode
  `\&12\catcode `\#12\catcode `\^12\catcode `\_12\catcode `\%12\relax}%
\providecommand \@@startlink[1]{}%
\providecommand \@@endlink[0]{}%
\providecommand \url  [0]{\begingroup\@sanitize@url \@url }%
\providecommand \@url [1]{\endgroup\@href {#1}{\urlprefix }}%
\providecommand \urlprefix  [0]{URL }%
\providecommand \Eprint [0]{\href }%
\providecommand \doibase [0]{https://doi.org/}%
\providecommand \selectlanguage [0]{\@gobble}%
\providecommand \bibinfo  [0]{\@secondoftwo}%
\providecommand \bibfield  [0]{\@secondoftwo}%
\providecommand \translation [1]{[#1]}%
\providecommand \BibitemOpen [0]{}%
\providecommand \bibitemStop [0]{}%
\providecommand \bibitemNoStop [0]{.\EOS\space}%
\providecommand \EOS [0]{\spacefactor3000\relax}%
\providecommand \BibitemShut  [1]{\csname bibitem#1\endcsname}%
\let\auto@bib@innerbib\@empty
\bibitem [{\citenamefont {Sauter}(1931)}]{sauter1931verhalten}%
  \BibitemOpen
  \bibfield  {author} {\bibinfo {author} {\bibfnamefont {F.}~\bibnamefont
  {Sauter}},\ }\href@noop {} {\bibfield  {journal} {\bibinfo  {journal}
  {Zeitschrift f{\"u}r Physik}\ }\textbf {\bibinfo {volume} {69}},\ \bibinfo
  {pages} {742} (\bibinfo {year} {1931})}\BibitemShut {NoStop}%
\bibitem [{\citenamefont {Heisenberg}\ and\ \citenamefont
  {Euler}(1936)}]{heisenberg1936folgerungen}%
  \BibitemOpen
  \bibfield  {author} {\bibinfo {author} {\bibfnamefont {W.}~\bibnamefont
  {Heisenberg}}\ and\ \bibinfo {author} {\bibfnamefont {H.}~\bibnamefont
  {Euler}},\ }\href@noop {} {\bibfield  {journal} {\bibinfo  {journal}
  {Zeitschrift f{\"u}r Physik}\ }\textbf {\bibinfo {volume} {98}},\ \bibinfo
  {pages} {714} (\bibinfo {year} {1936})}\BibitemShut {NoStop}%
\bibitem [{\citenamefont {Schwinger}(1951)}]{schwinger1951gauge}%
  \BibitemOpen
  \bibfield  {author} {\bibinfo {author} {\bibfnamefont {J.}~\bibnamefont
  {Schwinger}},\ }\href@noop {} {\bibfield  {journal} {\bibinfo  {journal}
  {Physical Review}\ }\textbf {\bibinfo {volume} {82}},\ \bibinfo {pages} {664}
  (\bibinfo {year} {1951})}\BibitemShut {NoStop}%
\bibitem [{\citenamefont {Greiner}\ \emph {et~al.}(2012)\citenamefont
  {Greiner}, \citenamefont {M{\"u}ller},\ and\ \citenamefont
  {Rafelski}}]{greiner2012quantum}%
  \BibitemOpen
  \bibfield  {author} {\bibinfo {author} {\bibfnamefont {W.}~\bibnamefont
  {Greiner}}, \bibinfo {author} {\bibfnamefont {B.}~\bibnamefont
  {M{\"u}ller}},\ and\ \bibinfo {author} {\bibfnamefont {J.}~\bibnamefont
  {Rafelski}},\ }\href@noop {} {\emph {\bibinfo {title} {Quantum
  electrodynamics of strong fields}}}\ (\bibinfo  {publisher} {Springer Science
  \& Business Media},\ \bibinfo {year} {2012})\BibitemShut {NoStop}%
\bibitem [{\citenamefont {Ritus}(1985)}]{ritus1985quantum}%
  \BibitemOpen
  \bibfield  {author} {\bibinfo {author} {\bibfnamefont {V.}~\bibnamefont
  {Ritus}},\ }\href@noop {} {\bibfield  {journal} {\bibinfo  {journal} {J. Sov.
  Laser Res. (United States)}\ }\textbf {\bibinfo {volume} {6}} (\bibinfo
  {year} {1985})}\BibitemShut {NoStop}%
\bibitem [{\citenamefont {Chen}\ and\ \citenamefont
  {Pellegrini}(1999)}]{chen1999boiling}%
  \BibitemOpen
  \bibfield  {author} {\bibinfo {author} {\bibfnamefont {P.}~\bibnamefont
  {Chen}}\ and\ \bibinfo {author} {\bibfnamefont {C.}~\bibnamefont
  {Pellegrini}},\ }\href@noop {} {\bibfield  {journal} {\bibinfo  {journal}
  {Quantum Aspects of Beam Physics, edited by P. Chen (World Scientific,
  1999)}\ ,\ \bibinfo {pages} {571}} (\bibinfo {year} {1999})}\BibitemShut
  {NoStop}%
\bibitem [{\citenamefont {Dunne}(2008)}]{dunne2008new}%
  \BibitemOpen
  \bibfield  {author} {\bibinfo {author} {\bibfnamefont {G.~V.}\ \bibnamefont
  {Dunne}},\ }\href@noop {} {\bibfield  {journal} {\bibinfo  {journal} {arXiv
  preprint arXiv:0812.3163}\ } (\bibinfo {year} {2008})}\BibitemShut {NoStop}%
\bibitem [{\citenamefont {Di~Piazza}\ \emph {et~al.}(2012)\citenamefont
  {Di~Piazza}, \citenamefont {M{\"u}ller}, \citenamefont {Hatsagortsyan},\ and\
  \citenamefont {Keitel}}]{di2012extremely}%
  \BibitemOpen
  \bibfield  {author} {\bibinfo {author} {\bibfnamefont {A.}~\bibnamefont
  {Di~Piazza}}, \bibinfo {author} {\bibfnamefont {C.}~\bibnamefont
  {M{\"u}ller}}, \bibinfo {author} {\bibfnamefont {K.}~\bibnamefont
  {Hatsagortsyan}},\ and\ \bibinfo {author} {\bibfnamefont {C.~H.}\
  \bibnamefont {Keitel}},\ }\href@noop {} {\bibfield  {journal} {\bibinfo
  {journal} {Reviews of Modern Physics}\ }\textbf {\bibinfo {volume} {84}},\
  \bibinfo {pages} {1177} (\bibinfo {year} {2012})}\BibitemShut {NoStop}%
\bibitem [{\citenamefont {Salgado}\ \emph {et~al.}(2021)\citenamefont
  {Salgado}, \citenamefont {Grafenstein}, \citenamefont {Golub}, \citenamefont
  {D{\"o}pp}, \citenamefont {Eckey}, \citenamefont {Hollatz}, \citenamefont
  {M{\"u}ller}, \citenamefont {Seidel}, \citenamefont {Seipt}, \citenamefont
  {Karsch} \emph {et~al.}}]{salgado2021towards}%
  \BibitemOpen
  \bibfield  {author} {\bibinfo {author} {\bibfnamefont {F.}~\bibnamefont
  {Salgado}}, \bibinfo {author} {\bibfnamefont {K.}~\bibnamefont
  {Grafenstein}}, \bibinfo {author} {\bibfnamefont {A.}~\bibnamefont {Golub}},
  \bibinfo {author} {\bibfnamefont {A.}~\bibnamefont {D{\"o}pp}}, \bibinfo
  {author} {\bibfnamefont {A.}~\bibnamefont {Eckey}}, \bibinfo {author}
  {\bibfnamefont {D.}~\bibnamefont {Hollatz}}, \bibinfo {author} {\bibfnamefont
  {C.}~\bibnamefont {M{\"u}ller}}, \bibinfo {author} {\bibfnamefont
  {A.}~\bibnamefont {Seidel}}, \bibinfo {author} {\bibfnamefont
  {D.}~\bibnamefont {Seipt}}, \bibinfo {author} {\bibfnamefont
  {S.}~\bibnamefont {Karsch}}, \emph {et~al.},\ }\href@noop {} {\bibfield
  {journal} {\bibinfo  {journal} {New Journal of Physics}\ }\textbf {\bibinfo
  {volume} {23}},\ \bibinfo {pages} {105002} (\bibinfo {year}
  {2021})}\BibitemShut {NoStop}%
\bibitem [{\citenamefont {Fedotov}\ \emph {et~al.}(2022)\citenamefont
  {Fedotov}, \citenamefont {Ilderton}, \citenamefont {Karbstein}, \citenamefont
  {King}, \citenamefont {Seipt}, \citenamefont {Taya},\ and\ \citenamefont
  {Torgrimsson}}]{fedotov2022advances}%
  \BibitemOpen
  \bibfield  {author} {\bibinfo {author} {\bibfnamefont {A.}~\bibnamefont
  {Fedotov}}, \bibinfo {author} {\bibfnamefont {A.}~\bibnamefont {Ilderton}},
  \bibinfo {author} {\bibfnamefont {F.}~\bibnamefont {Karbstein}}, \bibinfo
  {author} {\bibfnamefont {B.}~\bibnamefont {King}}, \bibinfo {author}
  {\bibfnamefont {D.}~\bibnamefont {Seipt}}, \bibinfo {author} {\bibfnamefont
  {H.}~\bibnamefont {Taya}},\ and\ \bibinfo {author} {\bibfnamefont
  {G.}~\bibnamefont {Torgrimsson}},\ }\href@noop {} {\bibfield  {journal}
  {\bibinfo  {journal} {arXiv preprint arXiv:2203.00019}\ } (\bibinfo {year}
  {2022})}\BibitemShut {NoStop}%
\bibitem [{\citenamefont {Andersson}(1998)}]{andersson1998lund}%
  \BibitemOpen
  \bibfield  {author} {\bibinfo {author} {\bibfnamefont {B.}~\bibnamefont
  {Andersson}},\ }\href@noop {} {\emph {\bibinfo {title} {The lund model}}},\
  \bibinfo {number} {7}\ (\bibinfo  {publisher} {Cambridge University Press},\
  \bibinfo {year} {1998})\BibitemShut {NoStop}%
\bibitem [{\citenamefont {Seiberg}\ and\ \citenamefont
  {Witten}(1999)}]{seiberg1999d1}%
  \BibitemOpen
  \bibfield  {author} {\bibinfo {author} {\bibfnamefont {N.}~\bibnamefont
  {Seiberg}}\ and\ \bibinfo {author} {\bibfnamefont {E.}~\bibnamefont
  {Witten}},\ }\href@noop {} {\bibfield  {journal} {\bibinfo  {journal}
  {Journal of High Energy Physics}\ }\textbf {\bibinfo {volume} {1999}},\
  \bibinfo {pages} {017} (\bibinfo {year} {1999})}\BibitemShut {NoStop}%
\bibitem [{\citenamefont {Ong}\ and\ \citenamefont {Chen}(2015)}]{Ong:2013bia}%
  \BibitemOpen
  \bibfield  {author} {\bibinfo {author} {\bibfnamefont {Y.~C.}\ \bibnamefont
  {Ong}}\ and\ \bibinfo {author} {\bibfnamefont {P.}~\bibnamefont {Chen}},\
  }in\ \href {https://doi.org/10.1142/9789814623995_0189} {\emph {\bibinfo
  {booktitle} {{13th Marcel Grossmann Meeting}}}}\ (\bibinfo {year} {2015})\
  pp.\ \bibinfo {pages} {1416--1418},\ \Eprint
  {https://arxiv.org/abs/1302.7162} {arXiv:1302.7162 [hep-th]} \BibitemShut
  {NoStop}%
\bibitem [{\citenamefont {Br{\'e}zin}\ and\ \citenamefont
  {Itzykson}(1970)}]{brezin1970pair}%
  \BibitemOpen
  \bibfield  {author} {\bibinfo {author} {\bibfnamefont {E.}~\bibnamefont
  {Br{\'e}zin}}\ and\ \bibinfo {author} {\bibfnamefont {C.}~\bibnamefont
  {Itzykson}},\ }\href@noop {} {\bibfield  {journal} {\bibinfo  {journal}
  {Physical Review D}\ }\textbf {\bibinfo {volume} {2}},\ \bibinfo {pages}
  {1191} (\bibinfo {year} {1970})}\BibitemShut {NoStop}%
\bibitem [{\citenamefont {Nikishov}\ and\ \citenamefont
  {Ritus}(1964)}]{nikishov1964quantum}%
  \BibitemOpen
  \bibfield  {author} {\bibinfo {author} {\bibfnamefont {A.}~\bibnamefont
  {Nikishov}}\ and\ \bibinfo {author} {\bibfnamefont {V.}~\bibnamefont
  {Ritus}},\ }\href@noop {} {\bibfield  {journal} {\bibinfo  {journal} {Sov.
  Phys. JETP}\ }\textbf {\bibinfo {volume} {19}},\ \bibinfo {pages} {529}
  (\bibinfo {year} {1964})}\BibitemShut {NoStop}%
\bibitem [{\citenamefont {Narozhnyi}\ \emph {et~al.}(1964)\citenamefont
  {Narozhnyi}, \citenamefont {Nikishov},\ and\ \citenamefont
  {Ritus}}]{narozhnyi1964quantum}%
  \BibitemOpen
  \bibfield  {author} {\bibinfo {author} {\bibfnamefont {N.}~\bibnamefont
  {Narozhnyi}}, \bibinfo {author} {\bibfnamefont {A.~I.}\ \bibnamefont
  {Nikishov}},\ and\ \bibinfo {author} {\bibfnamefont {V.}~\bibnamefont
  {Ritus}},\ }\href@noop {} {\bibfield  {journal} {\bibinfo  {journal} {Zh.
  Eksperim. i Teor. Fiz.}\ }\textbf {\bibinfo {volume} {47}} (\bibinfo {year}
  {1964})}\BibitemShut {NoStop}%
\bibitem [{\citenamefont {Nikishov}\ and\ \citenamefont
  {Ritus}(1965)}]{nikishov1965nonlinear}%
  \BibitemOpen
  \bibfield  {author} {\bibinfo {author} {\bibfnamefont {A.}~\bibnamefont
  {Nikishov}}\ and\ \bibinfo {author} {\bibfnamefont {V.}~\bibnamefont
  {Ritus}},\ }\href@noop {} {\bibfield  {journal} {\bibinfo  {journal} {Sov.
  Phys. JETP}\ }\textbf {\bibinfo {volume} {20}},\ \bibinfo {pages} {757}
  (\bibinfo {year} {1965})}\BibitemShut {NoStop}%
\bibitem [{\citenamefont {Nikishov}\ and\ \citenamefont
  {Ritus}(1967)}]{nikishov1967pair}%
  \BibitemOpen
  \bibfield  {author} {\bibinfo {author} {\bibfnamefont {A.}~\bibnamefont
  {Nikishov}}\ and\ \bibinfo {author} {\bibfnamefont {V.}~\bibnamefont
  {Ritus}},\ }\href@noop {} {\bibfield  {journal} {\bibinfo  {journal} {Sov.
  Phys. JETP}\ }\textbf {\bibinfo {volume} {25}},\ \bibinfo {pages} {1135}
  (\bibinfo {year} {1967})}\BibitemShut {NoStop}%
\bibitem [{\citenamefont {Bula}\ \emph {et~al.}(1996)\citenamefont {Bula},
  \citenamefont {McDonald}, \citenamefont {Prebys}, \citenamefont {Bamber},
  \citenamefont {Boege}, \citenamefont {Kotseroglou}, \citenamefont
  {Melissinos}, \citenamefont {Meyerhofer}, \citenamefont {Ragg}, \citenamefont
  {Burke} \emph {et~al.}}]{bula1996observation}%
  \BibitemOpen
  \bibfield  {author} {\bibinfo {author} {\bibfnamefont {C.}~\bibnamefont
  {Bula}}, \bibinfo {author} {\bibfnamefont {K.}~\bibnamefont {McDonald}},
  \bibinfo {author} {\bibfnamefont {E.}~\bibnamefont {Prebys}}, \bibinfo
  {author} {\bibfnamefont {C.}~\bibnamefont {Bamber}}, \bibinfo {author}
  {\bibfnamefont {S.}~\bibnamefont {Boege}}, \bibinfo {author} {\bibfnamefont
  {T.}~\bibnamefont {Kotseroglou}}, \bibinfo {author} {\bibfnamefont
  {A.}~\bibnamefont {Melissinos}}, \bibinfo {author} {\bibfnamefont
  {D.}~\bibnamefont {Meyerhofer}}, \bibinfo {author} {\bibfnamefont
  {W.}~\bibnamefont {Ragg}}, \bibinfo {author} {\bibfnamefont {D.}~\bibnamefont
  {Burke}}, \emph {et~al.},\ }\href@noop {} {\bibfield  {journal} {\bibinfo
  {journal} {Physical Review Letters}\ }\textbf {\bibinfo {volume} {76}},\
  \bibinfo {pages} {3116} (\bibinfo {year} {1996})}\BibitemShut {NoStop}%
\bibitem [{\citenamefont {Bamber}\ \emph {et~al.}(1999)\citenamefont {Bamber},
  \citenamefont {Boege}, \citenamefont {Koffas}, \citenamefont {Kotseroglou},
  \citenamefont {Melissinos}, \citenamefont {Meyerhofer}, \citenamefont {Reis},
  \citenamefont {Ragg}, \citenamefont {Bula}, \citenamefont {McDonald} \emph
  {et~al.}}]{bamber1999studies}%
  \BibitemOpen
  \bibfield  {author} {\bibinfo {author} {\bibfnamefont {C.}~\bibnamefont
  {Bamber}}, \bibinfo {author} {\bibfnamefont {S.}~\bibnamefont {Boege}},
  \bibinfo {author} {\bibfnamefont {T.}~\bibnamefont {Koffas}}, \bibinfo
  {author} {\bibfnamefont {T.}~\bibnamefont {Kotseroglou}}, \bibinfo {author}
  {\bibfnamefont {A.}~\bibnamefont {Melissinos}}, \bibinfo {author}
  {\bibfnamefont {D.}~\bibnamefont {Meyerhofer}}, \bibinfo {author}
  {\bibfnamefont {D.}~\bibnamefont {Reis}}, \bibinfo {author} {\bibfnamefont
  {W.}~\bibnamefont {Ragg}}, \bibinfo {author} {\bibfnamefont {C.}~\bibnamefont
  {Bula}}, \bibinfo {author} {\bibfnamefont {K.}~\bibnamefont {McDonald}},
  \emph {et~al.},\ }\href@noop {} {\bibfield  {journal} {\bibinfo  {journal}
  {Physical Review D}\ }\textbf {\bibinfo {volume} {60}},\ \bibinfo {pages}
  {092004} (\bibinfo {year} {1999})}\BibitemShut {NoStop}%
\bibitem [{\citenamefont {Mourou}\ \emph {et~al.}(2011)\citenamefont {Mourou},
  \citenamefont {Korn}, \citenamefont {Sandner},\ and\ \citenamefont
  {Collier}}]{eliwhite}%
  \BibitemOpen
  \bibfield  {author} {\bibinfo {author} {\bibfnamefont {G.}~\bibnamefont
  {Mourou}}, \bibinfo {author} {\bibfnamefont {G.}~\bibnamefont {Korn}},
  \bibinfo {author} {\bibfnamefont {W.}~\bibnamefont {Sandner}},\ and\ \bibinfo
  {author} {\bibfnamefont {J.~e.}\ \bibnamefont {Collier}},\ }\href
  {http://www.eli-laser.eu/} {\emph {\bibinfo {title} {ELI White Book}}}\
  (\bibinfo  {publisher} {Thoss Media, Berlin},\ \bibinfo {year}
  {2011})\BibitemShut {NoStop}%
\bibitem [{\citenamefont {Nees}\ \emph {et~al.}(2020)\citenamefont {Nees},
  \citenamefont {Maksimchuk}, \citenamefont {Kalinchenko}, \citenamefont {Hou},
  \citenamefont {Ma}, \citenamefont {Campbell}, \citenamefont {McKelvey},
  \citenamefont {Willingale}, \citenamefont {Jovanovic}, \citenamefont {Kuranz}
  \emph {et~al.}}]{nees2020zeus}%
  \BibitemOpen
  \bibfield  {author} {\bibinfo {author} {\bibfnamefont {J.}~\bibnamefont
  {Nees}}, \bibinfo {author} {\bibfnamefont {A.}~\bibnamefont {Maksimchuk}},
  \bibinfo {author} {\bibfnamefont {G.}~\bibnamefont {Kalinchenko}}, \bibinfo
  {author} {\bibfnamefont {B.}~\bibnamefont {Hou}}, \bibinfo {author}
  {\bibfnamefont {Y.}~\bibnamefont {Ma}}, \bibinfo {author} {\bibfnamefont
  {P.}~\bibnamefont {Campbell}}, \bibinfo {author} {\bibfnamefont
  {A.}~\bibnamefont {McKelvey}}, \bibinfo {author} {\bibfnamefont
  {L.}~\bibnamefont {Willingale}}, \bibinfo {author} {\bibfnamefont
  {I.}~\bibnamefont {Jovanovic}}, \bibinfo {author} {\bibfnamefont
  {C.}~\bibnamefont {Kuranz}}, \emph {et~al.},\ }in\ \href@noop {} {\emph
  {\bibinfo {booktitle} {2020 Conference on Lasers and Electro-Optics
  (CLEO)}}}\ (\bibinfo {organization} {IEEE},\ \bibinfo {year} {2020})\ pp.\
  \bibinfo {pages} {1--2}\BibitemShut {NoStop}%
\bibitem [{\citenamefont {Labun}\ and\ \citenamefont
  {Rafelski}(2009)}]{labun2009vacuum}%
  \BibitemOpen
  \bibfield  {author} {\bibinfo {author} {\bibfnamefont {L.}~\bibnamefont
  {Labun}}\ and\ \bibinfo {author} {\bibfnamefont {J.}~\bibnamefont
  {Rafelski}},\ }\href@noop {} {\bibfield  {journal} {\bibinfo  {journal}
  {Physical Review D}\ }\textbf {\bibinfo {volume} {79}},\ \bibinfo {pages}
  {057901} (\bibinfo {year} {2009})}\BibitemShut {NoStop}%
\bibitem [{\citenamefont {Dittrich}\ and\ \citenamefont
  {Gies}(2000)}]{dittrich2000probing}%
  \BibitemOpen
  \bibfield  {author} {\bibinfo {author} {\bibfnamefont {W.}~\bibnamefont
  {Dittrich}}\ and\ \bibinfo {author} {\bibfnamefont {H.}~\bibnamefont
  {Gies}},\ }\href@noop {} {\emph {\bibinfo {title} {Probing the quantum
  vacuum}}},\ Vol.\ \bibinfo {volume} {166}\ (\bibinfo  {publisher} {Springer
  Science \& Business Media},\ \bibinfo {year} {2000})\BibitemShut {NoStop}%
\bibitem [{\citenamefont {Dunne}\ \emph {et~al.}(2009)\citenamefont {Dunne},
  \citenamefont {Gies},\ and\ \citenamefont
  {Sch{\"u}tzhold}}]{dunne2009catalysis}%
  \BibitemOpen
  \bibfield  {author} {\bibinfo {author} {\bibfnamefont {G.~V.}\ \bibnamefont
  {Dunne}}, \bibinfo {author} {\bibfnamefont {H.}~\bibnamefont {Gies}},\ and\
  \bibinfo {author} {\bibfnamefont {R.}~\bibnamefont {Sch{\"u}tzhold}},\
  }\href@noop {} {\bibfield  {journal} {\bibinfo  {journal} {Physical Review
  D}\ }\textbf {\bibinfo {volume} {80}},\ \bibinfo {pages} {111301} (\bibinfo
  {year} {2009})}\BibitemShut {NoStop}%
\bibitem [{\citenamefont {Krajewska}\ and\ \citenamefont
  {Kami{\'n}ski}(2012)}]{krajewska2012breit}%
  \BibitemOpen
  \bibfield  {author} {\bibinfo {author} {\bibfnamefont {K.}~\bibnamefont
  {Krajewska}}\ and\ \bibinfo {author} {\bibfnamefont {J.}~\bibnamefont
  {Kami{\'n}ski}},\ }\href@noop {} {\bibfield  {journal} {\bibinfo  {journal}
  {Physical Review A}\ }\textbf {\bibinfo {volume} {86}},\ \bibinfo {pages}
  {052104} (\bibinfo {year} {2012})}\BibitemShut {NoStop}%
\bibitem [{\citenamefont {Nousch}\ \emph {et~al.}(2012)\citenamefont {Nousch},
  \citenamefont {Seipt}, \citenamefont {K{\"a}mpfer},\ and\ \citenamefont
  {Titov}}]{nousch2012pair}%
  \BibitemOpen
  \bibfield  {author} {\bibinfo {author} {\bibfnamefont {T.}~\bibnamefont
  {Nousch}}, \bibinfo {author} {\bibfnamefont {D.}~\bibnamefont {Seipt}},
  \bibinfo {author} {\bibfnamefont {B.}~\bibnamefont {K{\"a}mpfer}},\ and\
  \bibinfo {author} {\bibfnamefont {A.}~\bibnamefont {Titov}},\ }\href@noop {}
  {\bibfield  {journal} {\bibinfo  {journal} {Physics Letters B}\ }\textbf
  {\bibinfo {volume} {715}},\ \bibinfo {pages} {246} (\bibinfo {year}
  {2012})}\BibitemShut {NoStop}%
\bibitem [{\citenamefont {Titov}\ \emph {et~al.}(2012)\citenamefont {Titov},
  \citenamefont {Takabe}, \citenamefont {Kampfer},\ and\ \citenamefont
  {Hosaka}}]{Titov:2012rd}%
  \BibitemOpen
  \bibfield  {author} {\bibinfo {author} {\bibfnamefont {A.~I.}\ \bibnamefont
  {Titov}}, \bibinfo {author} {\bibfnamefont {H.}~\bibnamefont {Takabe}},
  \bibinfo {author} {\bibfnamefont {B.}~\bibnamefont {Kampfer}},\ and\ \bibinfo
  {author} {\bibfnamefont {A.}~\bibnamefont {Hosaka}},\ }\href
  {https://doi.org/10.1103/PhysRevLett.108.240406} {\bibfield  {journal}
  {\bibinfo  {journal} {Phys. Rev. Lett.}\ }\textbf {\bibinfo {volume} {108}},\
  \bibinfo {pages} {240406} (\bibinfo {year} {2012})},\ \Eprint
  {https://arxiv.org/abs/1205.3880} {arXiv:1205.3880 [hep-ph]} \BibitemShut
  {NoStop}%
\bibitem [{\citenamefont {Karbstein}(2013)}]{karbstein2013photon}%
  \BibitemOpen
  \bibfield  {author} {\bibinfo {author} {\bibfnamefont {F.}~\bibnamefont
  {Karbstein}},\ }\href@noop {} {\bibfield  {journal} {\bibinfo  {journal}
  {Physical Review D}\ }\textbf {\bibinfo {volume} {88}},\ \bibinfo {pages}
  {085033} (\bibinfo {year} {2013})}\BibitemShut {NoStop}%
\bibitem [{\citenamefont {Titov}\ \emph {et~al.}(2013)\citenamefont {Titov},
  \citenamefont {K{\"a}mpfer}, \citenamefont {Takabe},\ and\ \citenamefont
  {Hosaka}}]{titov2013breit}%
  \BibitemOpen
  \bibfield  {author} {\bibinfo {author} {\bibfnamefont {A.}~\bibnamefont
  {Titov}}, \bibinfo {author} {\bibfnamefont {B.}~\bibnamefont {K{\"a}mpfer}},
  \bibinfo {author} {\bibfnamefont {H.}~\bibnamefont {Takabe}},\ and\ \bibinfo
  {author} {\bibfnamefont {A.}~\bibnamefont {Hosaka}},\ }\href@noop {}
  {\bibfield  {journal} {\bibinfo  {journal} {Physical Review A}\ }\textbf
  {\bibinfo {volume} {87}},\ \bibinfo {pages} {042106} (\bibinfo {year}
  {2013})}\BibitemShut {NoStop}%
\bibitem [{\citenamefont {Wu}\ and\ \citenamefont
  {Xue}(2014)}]{wu2014nonlinear}%
  \BibitemOpen
  \bibfield  {author} {\bibinfo {author} {\bibfnamefont {Y.-B.}\ \bibnamefont
  {Wu}}\ and\ \bibinfo {author} {\bibfnamefont {S.-S.}\ \bibnamefont {Xue}},\
  }\href@noop {} {\bibfield  {journal} {\bibinfo  {journal} {Physical Review
  D}\ }\textbf {\bibinfo {volume} {90}},\ \bibinfo {pages} {013009} (\bibinfo
  {year} {2014})}\BibitemShut {NoStop}%
\bibitem [{\citenamefont {Meuren}\ \emph {et~al.}(2015)\citenamefont {Meuren},
  \citenamefont {Hatsagortsyan}, \citenamefont {Keitel},\ and\ \citenamefont
  {Di~Piazza}}]{meuren2015polarization}%
  \BibitemOpen
  \bibfield  {author} {\bibinfo {author} {\bibfnamefont {S.}~\bibnamefont
  {Meuren}}, \bibinfo {author} {\bibfnamefont {K.~Z.}\ \bibnamefont
  {Hatsagortsyan}}, \bibinfo {author} {\bibfnamefont {C.~H.}\ \bibnamefont
  {Keitel}},\ and\ \bibinfo {author} {\bibfnamefont {A.}~\bibnamefont
  {Di~Piazza}},\ }\href@noop {} {\bibfield  {journal} {\bibinfo  {journal}
  {Physical Review D}\ }\textbf {\bibinfo {volume} {91}},\ \bibinfo {pages}
  {013009} (\bibinfo {year} {2015})}\BibitemShut {NoStop}%
\bibitem [{\citenamefont {Di~Piazza}(2016)}]{di2016nonlinear}%
  \BibitemOpen
  \bibfield  {author} {\bibinfo {author} {\bibfnamefont {A.}~\bibnamefont
  {Di~Piazza}},\ }\href@noop {} {\bibfield  {journal} {\bibinfo  {journal}
  {Physical Review Letters}\ }\textbf {\bibinfo {volume} {117}},\ \bibinfo
  {pages} {213201} (\bibinfo {year} {2016})}\BibitemShut {NoStop}%
\bibitem [{\citenamefont {Otto}\ \emph {et~al.}(2016)\citenamefont {Otto},
  \citenamefont {Nousch}, \citenamefont {Seipt}, \citenamefont {K{\"a}mpfer},
  \citenamefont {Blaschke}, \citenamefont {Panferov}, \citenamefont
  {Smolyansky},\ and\ \citenamefont {Titov}}]{otto2016pair}%
  \BibitemOpen
  \bibfield  {author} {\bibinfo {author} {\bibfnamefont {A.}~\bibnamefont
  {Otto}}, \bibinfo {author} {\bibfnamefont {T.}~\bibnamefont {Nousch}},
  \bibinfo {author} {\bibfnamefont {D.}~\bibnamefont {Seipt}}, \bibinfo
  {author} {\bibfnamefont {B.}~\bibnamefont {K{\"a}mpfer}}, \bibinfo {author}
  {\bibfnamefont {D.}~\bibnamefont {Blaschke}}, \bibinfo {author}
  {\bibfnamefont {A.}~\bibnamefont {Panferov}}, \bibinfo {author}
  {\bibfnamefont {S.}~\bibnamefont {Smolyansky}},\ and\ \bibinfo {author}
  {\bibfnamefont {A.}~\bibnamefont {Titov}},\ }\href@noop {} {\bibfield
  {journal} {\bibinfo  {journal} {Journal of Plasma Physics}\ }\textbf
  {\bibinfo {volume} {82}},\ \bibinfo {pages} {655820301} (\bibinfo {year}
  {2016})}\BibitemShut {NoStop}%
\bibitem [{\citenamefont {Blackburn}\ and\ \citenamefont
  {Marklund}(2018)}]{blackburn2018nonlinear}%
  \BibitemOpen
  \bibfield  {author} {\bibinfo {author} {\bibfnamefont {T.}~\bibnamefont
  {Blackburn}}\ and\ \bibinfo {author} {\bibfnamefont {M.}~\bibnamefont
  {Marklund}},\ }\href@noop {} {\bibfield  {journal} {\bibinfo  {journal}
  {Plasma Physics and Controlled Fusion}\ }\textbf {\bibinfo {volume} {60}},\
  \bibinfo {pages} {054009} (\bibinfo {year} {2018})}\BibitemShut {NoStop}%
\bibitem [{\citenamefont {Mackenroth}\ and\ \citenamefont
  {Di~Piazza}(2018)}]{mackenroth2018nonlinear}%
  \BibitemOpen
  \bibfield  {author} {\bibinfo {author} {\bibfnamefont {F.}~\bibnamefont
  {Mackenroth}}\ and\ \bibinfo {author} {\bibfnamefont {A.}~\bibnamefont
  {Di~Piazza}},\ }\href@noop {} {\bibfield  {journal} {\bibinfo  {journal}
  {Physical Review D}\ }\textbf {\bibinfo {volume} {98}},\ \bibinfo {pages}
  {116002} (\bibinfo {year} {2018})}\BibitemShut {NoStop}%
\bibitem [{\citenamefont {King}(2020)}]{king2020uniform}%
  \BibitemOpen
  \bibfield  {author} {\bibinfo {author} {\bibfnamefont {B.}~\bibnamefont
  {King}},\ }\href@noop {} {\bibfield  {journal} {\bibinfo  {journal} {Physical
  Review A}\ }\textbf {\bibinfo {volume} {101}},\ \bibinfo {pages} {042508}
  (\bibinfo {year} {2020})}\BibitemShut {NoStop}%
\bibitem [{\citenamefont {Seipt}\ and\ \citenamefont
  {King}(2020)}]{seipt2020spin}%
  \BibitemOpen
  \bibfield  {author} {\bibinfo {author} {\bibfnamefont {D.}~\bibnamefont
  {Seipt}}\ and\ \bibinfo {author} {\bibfnamefont {B.}~\bibnamefont {King}},\
  }\href@noop {} {\bibfield  {journal} {\bibinfo  {journal} {Physical Review
  A}\ }\textbf {\bibinfo {volume} {102}},\ \bibinfo {pages} {052805} (\bibinfo
  {year} {2020})}\BibitemShut {NoStop}%
\bibitem [{\citenamefont {Wan}\ \emph {et~al.}(2020)\citenamefont {Wan},
  \citenamefont {Wang}, \citenamefont {Guo}, \citenamefont {Chen},
  \citenamefont {Shaisultanov}, \citenamefont {Xu}, \citenamefont
  {Hatsagortsyan}, \citenamefont {Keitel},\ and\ \citenamefont
  {Li}}]{wan2020high}%
  \BibitemOpen
  \bibfield  {author} {\bibinfo {author} {\bibfnamefont {F.}~\bibnamefont
  {Wan}}, \bibinfo {author} {\bibfnamefont {Y.}~\bibnamefont {Wang}}, \bibinfo
  {author} {\bibfnamefont {R.-T.}\ \bibnamefont {Guo}}, \bibinfo {author}
  {\bibfnamefont {Y.-Y.}\ \bibnamefont {Chen}}, \bibinfo {author}
  {\bibfnamefont {R.}~\bibnamefont {Shaisultanov}}, \bibinfo {author}
  {\bibfnamefont {Z.-F.}\ \bibnamefont {Xu}}, \bibinfo {author} {\bibfnamefont
  {K.~Z.}\ \bibnamefont {Hatsagortsyan}}, \bibinfo {author} {\bibfnamefont
  {C.~H.}\ \bibnamefont {Keitel}},\ and\ \bibinfo {author} {\bibfnamefont
  {J.-X.}\ \bibnamefont {Li}},\ }\href@noop {} {\bibfield  {journal} {\bibinfo
  {journal} {Physical Review Research}\ }\textbf {\bibinfo {volume} {2}},\
  \bibinfo {pages} {032049} (\bibinfo {year} {2020})}\BibitemShut {NoStop}%
\bibitem [{\citenamefont {Mercuri-Baron}\ \emph {et~al.}(2021)\citenamefont
  {Mercuri-Baron}, \citenamefont {Grech}, \citenamefont {Niel}, \citenamefont
  {Grassi}, \citenamefont {Lobet}, \citenamefont {Di~Piazza},\ and\
  \citenamefont {Riconda}}]{mercuri2021impact}%
  \BibitemOpen
  \bibfield  {author} {\bibinfo {author} {\bibfnamefont {A.}~\bibnamefont
  {Mercuri-Baron}}, \bibinfo {author} {\bibfnamefont {M.}~\bibnamefont
  {Grech}}, \bibinfo {author} {\bibfnamefont {F.}~\bibnamefont {Niel}},
  \bibinfo {author} {\bibfnamefont {A.}~\bibnamefont {Grassi}}, \bibinfo
  {author} {\bibfnamefont {M.}~\bibnamefont {Lobet}}, \bibinfo {author}
  {\bibfnamefont {A.}~\bibnamefont {Di~Piazza}},\ and\ \bibinfo {author}
  {\bibfnamefont {C.}~\bibnamefont {Riconda}},\ }\href@noop {} {\bibfield
  {journal} {\bibinfo  {journal} {New Journal of Physics}\ }\textbf {\bibinfo
  {volume} {23}},\ \bibinfo {pages} {085006} (\bibinfo {year}
  {2021})}\BibitemShut {NoStop}%
\bibitem [{\citenamefont {Song}\ \emph {et~al.}(2021)\citenamefont {Song},
  \citenamefont {Wang}, \citenamefont {Li}, \citenamefont {Li}, \citenamefont
  {Li}, \citenamefont {Sheng}, \citenamefont {Chen},\ and\ \citenamefont
  {Zhang}}]{song2021spin}%
  \BibitemOpen
  \bibfield  {author} {\bibinfo {author} {\bibfnamefont {H.-H.}\ \bibnamefont
  {Song}}, \bibinfo {author} {\bibfnamefont {W.-M.}\ \bibnamefont {Wang}},
  \bibinfo {author} {\bibfnamefont {Y.-F.}\ \bibnamefont {Li}}, \bibinfo
  {author} {\bibfnamefont {B.-J.}\ \bibnamefont {Li}}, \bibinfo {author}
  {\bibfnamefont {Y.-T.}\ \bibnamefont {Li}}, \bibinfo {author} {\bibfnamefont
  {Z.-M.}\ \bibnamefont {Sheng}}, \bibinfo {author} {\bibfnamefont {L.-M.}\
  \bibnamefont {Chen}},\ and\ \bibinfo {author} {\bibfnamefont
  {J.}~\bibnamefont {Zhang}},\ }\href@noop {} {\bibfield  {journal} {\bibinfo
  {journal} {New Journal of Physics}\ }\textbf {\bibinfo {volume} {23}},\
  \bibinfo {pages} {075005} (\bibinfo {year} {2021})}\BibitemShut {NoStop}%
\bibitem [{\citenamefont {Tang}\ and\ \citenamefont
  {King}(2021)}]{tang2021pulse}%
  \BibitemOpen
  \bibfield  {author} {\bibinfo {author} {\bibfnamefont {S.}~\bibnamefont
  {Tang}}\ and\ \bibinfo {author} {\bibfnamefont {B.}~\bibnamefont {King}},\
  }\href@noop {} {\bibfield  {journal} {\bibinfo  {journal} {Physical Review
  D}\ }\textbf {\bibinfo {volume} {104}},\ \bibinfo {pages} {096019} (\bibinfo
  {year} {2021})}\BibitemShut {NoStop}%
\bibitem [{\citenamefont {Gao}\ and\ \citenamefont
  {Tang}(2022)}]{gao2022optimal}%
  \BibitemOpen
  \bibfield  {author} {\bibinfo {author} {\bibfnamefont {Y.}~\bibnamefont
  {Gao}}\ and\ \bibinfo {author} {\bibfnamefont {S.}~\bibnamefont {Tang}},\
  }\href@noop {} {\bibfield  {journal} {\bibinfo  {journal} {Physical Review
  D}\ }\textbf {\bibinfo {volume} {106}},\ \bibinfo {pages} {056003} (\bibinfo
  {year} {2022})}\BibitemShut {NoStop}%
\bibitem [{\citenamefont {Golub}\ \emph {et~al.}(2022)\citenamefont {Golub},
  \citenamefont {Villalba-Ch{\'a}vez},\ and\ \citenamefont
  {M{\"u}ller}}]{golub2022nonlinear}%
  \BibitemOpen
  \bibfield  {author} {\bibinfo {author} {\bibfnamefont {A.}~\bibnamefont
  {Golub}}, \bibinfo {author} {\bibfnamefont {S.}~\bibnamefont
  {Villalba-Ch{\'a}vez}},\ and\ \bibinfo {author} {\bibfnamefont
  {C.}~\bibnamefont {M{\"u}ller}},\ }\href@noop {} {\bibfield  {journal}
  {\bibinfo  {journal} {Physical Review D}\ }\textbf {\bibinfo {volume}
  {105}},\ \bibinfo {pages} {116016} (\bibinfo {year} {2022})}\BibitemShut
  {NoStop}%
\bibitem [{\citenamefont {Tang}(2022)}]{tang2022fully}%
  \BibitemOpen
  \bibfield  {author} {\bibinfo {author} {\bibfnamefont {S.}~\bibnamefont
  {Tang}},\ }\href@noop {} {\bibfield  {journal} {\bibinfo  {journal} {Physical
  Review D}\ }\textbf {\bibinfo {volume} {105}},\ \bibinfo {pages} {056018}
  (\bibinfo {year} {2022})}\BibitemShut {NoStop}%
\bibitem [{\citenamefont {MacLeod}\ \emph {et~al.}(2023)\citenamefont
  {MacLeod}, \citenamefont {Hadjisolomou}, \citenamefont {Jeong},\ and\
  \citenamefont {Bulanov}}]{macleod2023all}%
  \BibitemOpen
  \bibfield  {author} {\bibinfo {author} {\bibfnamefont {A.~J.}\ \bibnamefont
  {MacLeod}}, \bibinfo {author} {\bibfnamefont {P.}~\bibnamefont
  {Hadjisolomou}}, \bibinfo {author} {\bibfnamefont {T.~M.}\ \bibnamefont
  {Jeong}},\ and\ \bibinfo {author} {\bibfnamefont {S.~V.}\ \bibnamefont
  {Bulanov}},\ }\href@noop {} {\bibfield  {journal} {\bibinfo  {journal}
  {Physical Review A}\ }\textbf {\bibinfo {volume} {107}},\ \bibinfo {pages}
  {012215} (\bibinfo {year} {2023})}\BibitemShut {NoStop}%
\bibitem [{\citenamefont {Chiu}\ and\ \citenamefont
  {Nussinov}(1979)}]{chiu1979diagrammatic}%
  \BibitemOpen
  \bibfield  {author} {\bibinfo {author} {\bibfnamefont {C.}~\bibnamefont
  {Chiu}}\ and\ \bibinfo {author} {\bibfnamefont {S.}~\bibnamefont
  {Nussinov}},\ }\href@noop {} {\bibfield  {journal} {\bibinfo  {journal}
  {Physical Review D}\ }\textbf {\bibinfo {volume} {20}},\ \bibinfo {pages}
  {945} (\bibinfo {year} {1979})}\BibitemShut {NoStop}%
\bibitem [{\citenamefont {Labun}\ and\ \citenamefont
  {Rafelski}(2011)}]{labun2011spectra}%
  \BibitemOpen
  \bibfield  {author} {\bibinfo {author} {\bibfnamefont {L.}~\bibnamefont
  {Labun}}\ and\ \bibinfo {author} {\bibfnamefont {J.}~\bibnamefont
  {Rafelski}},\ }\href@noop {} {\bibfield  {journal} {\bibinfo  {journal}
  {Physical Review D}\ }\textbf {\bibinfo {volume} {84}},\ \bibinfo {pages}
  {033003} (\bibinfo {year} {2011})}\BibitemShut {NoStop}%
\bibitem [{\citenamefont {Fedotov}\ \emph {et~al.}(2010)\citenamefont
  {Fedotov}, \citenamefont {Narozhny}, \citenamefont {Mourou},\ and\
  \citenamefont {Korn}}]{fedotov2010limitations}%
  \BibitemOpen
  \bibfield  {author} {\bibinfo {author} {\bibfnamefont {A.}~\bibnamefont
  {Fedotov}}, \bibinfo {author} {\bibfnamefont {N.}~\bibnamefont {Narozhny}},
  \bibinfo {author} {\bibfnamefont {G.}~\bibnamefont {Mourou}},\ and\ \bibinfo
  {author} {\bibfnamefont {G.}~\bibnamefont {Korn}},\ }\href@noop {} {\bibfield
   {journal} {\bibinfo  {journal} {Physical review letters}\ }\textbf {\bibinfo
  {volume} {105}},\ \bibinfo {pages} {080402} (\bibinfo {year}
  {2010})}\BibitemShut {NoStop}%
\bibitem [{\citenamefont {Elkina}\ \emph {et~al.}(2011)\citenamefont {Elkina},
  \citenamefont {Fedotov}, \citenamefont {Kostyukov}, \citenamefont {Legkov},
  \citenamefont {Narozhny}, \citenamefont {Nerush},\ and\ \citenamefont
  {Ruhl}}]{elkina2011qed}%
  \BibitemOpen
  \bibfield  {author} {\bibinfo {author} {\bibfnamefont {N.}~\bibnamefont
  {Elkina}}, \bibinfo {author} {\bibfnamefont {A.}~\bibnamefont {Fedotov}},
  \bibinfo {author} {\bibfnamefont {I.~Y.}\ \bibnamefont {Kostyukov}}, \bibinfo
  {author} {\bibfnamefont {M.}~\bibnamefont {Legkov}}, \bibinfo {author}
  {\bibfnamefont {N.}~\bibnamefont {Narozhny}}, \bibinfo {author}
  {\bibfnamefont {E.}~\bibnamefont {Nerush}},\ and\ \bibinfo {author}
  {\bibfnamefont {H.}~\bibnamefont {Ruhl}},\ }\href@noop {} {\bibfield
  {journal} {\bibinfo  {journal} {Physical Review Special Topics-Accelerators
  and Beams}\ }\textbf {\bibinfo {volume} {14}},\ \bibinfo {pages} {054401}
  (\bibinfo {year} {2011})}\BibitemShut {NoStop}%
\bibitem [{\citenamefont {Hebenstreit}\ \emph {et~al.}(2009)\citenamefont
  {Hebenstreit}, \citenamefont {Alkofer}, \citenamefont {Dunne},\ and\
  \citenamefont {Gies}}]{hebenstreit2009momentum}%
  \BibitemOpen
  \bibfield  {author} {\bibinfo {author} {\bibfnamefont {F.}~\bibnamefont
  {Hebenstreit}}, \bibinfo {author} {\bibfnamefont {R.}~\bibnamefont
  {Alkofer}}, \bibinfo {author} {\bibfnamefont {G.~V.}\ \bibnamefont {Dunne}},\
  and\ \bibinfo {author} {\bibfnamefont {H.}~\bibnamefont {Gies}},\ }\href@noop
  {} {\bibfield  {journal} {\bibinfo  {journal} {Physical review letters}\
  }\textbf {\bibinfo {volume} {102}},\ \bibinfo {pages} {150404} (\bibinfo
  {year} {2009})}\BibitemShut {NoStop}%
\bibitem [{\citenamefont {Orthaber}\ \emph {et~al.}(2011)\citenamefont
  {Orthaber}, \citenamefont {Hebenstreit},\ and\ \citenamefont
  {Alkofer}}]{orthaber2011momentum}%
  \BibitemOpen
  \bibfield  {author} {\bibinfo {author} {\bibfnamefont {M.}~\bibnamefont
  {Orthaber}}, \bibinfo {author} {\bibfnamefont {F.}~\bibnamefont
  {Hebenstreit}},\ and\ \bibinfo {author} {\bibfnamefont {R.}~\bibnamefont
  {Alkofer}},\ }\href@noop {} {\bibfield  {journal} {\bibinfo  {journal}
  {Physics Letters B}\ }\textbf {\bibinfo {volume} {698}},\ \bibinfo {pages}
  {80} (\bibinfo {year} {2011})}\BibitemShut {NoStop}%
\bibitem [{\citenamefont {Fey}\ and\ \citenamefont
  {Sch{\"u}tzhold}(2012)}]{fey2012momentum}%
  \BibitemOpen
  \bibfield  {author} {\bibinfo {author} {\bibfnamefont {C.}~\bibnamefont
  {Fey}}\ and\ \bibinfo {author} {\bibfnamefont {R.}~\bibnamefont
  {Sch{\"u}tzhold}},\ }\href@noop {} {\bibfield  {journal} {\bibinfo  {journal}
  {Physical Review D}\ }\textbf {\bibinfo {volume} {85}},\ \bibinfo {pages}
  {025004} (\bibinfo {year} {2012})}\BibitemShut {NoStop}%
\bibitem [{\citenamefont {Nikishov}(1970)}]{Nikishov:1970br}%
  \BibitemOpen
  \bibfield  {author} {\bibinfo {author} {\bibfnamefont {A.~I.}\ \bibnamefont
  {Nikishov}},\ }\href {https://doi.org/10.1016/0550-3213(70)90527-4}
  {\bibfield  {journal} {\bibinfo  {journal} {Nucl. Phys. B}\ }\textbf
  {\bibinfo {volume} {21}},\ \bibinfo {pages} {346} (\bibinfo {year}
  {1970})}\BibitemShut {NoStop}%
\bibitem [{\citenamefont {Cohen}\ and\ \citenamefont
  {McGady}(2008)}]{Cohen:2008wz}%
  \BibitemOpen
  \bibfield  {author} {\bibinfo {author} {\bibfnamefont {T.~D.}\ \bibnamefont
  {Cohen}}\ and\ \bibinfo {author} {\bibfnamefont {D.~A.}\ \bibnamefont
  {McGady}},\ }\href {https://doi.org/10.1103/PhysRevD.78.036008} {\bibfield
  {journal} {\bibinfo  {journal} {Phys. Rev. D}\ }\textbf {\bibinfo {volume}
  {78}},\ \bibinfo {pages} {036008} (\bibinfo {year} {2008})},\ \Eprint
  {https://arxiv.org/abs/0807.1117} {arXiv:0807.1117 [hep-ph]} \BibitemShut
  {NoStop}%
\bibitem [{\citenamefont {Baier}\ and\ \citenamefont
  {Katkov}(2005)}]{Baier:2003hf}%
  \BibitemOpen
  \bibfield  {author} {\bibinfo {author} {\bibfnamefont {V.~N.}\ \bibnamefont
  {Baier}}\ and\ \bibinfo {author} {\bibfnamefont {V.~M.}\ \bibnamefont
  {Katkov}},\ }\href {https://doi.org/10.1016/j.physrep.2004.11.003} {\bibfield
   {journal} {\bibinfo  {journal} {Phys. Rept.}\ }\textbf {\bibinfo {volume}
  {409}},\ \bibinfo {pages} {261} (\bibinfo {year} {2005})},\ \Eprint
  {https://arxiv.org/abs/hep-ph/0309211} {arXiv:hep-ph/0309211} \BibitemShut
  {NoStop}%
\bibitem [{\citenamefont {Landau}\ and\ \citenamefont
  {Lifshitz}(1971)}]{landau1971classical}%
  \BibitemOpen
  \bibfield  {author} {\bibinfo {author} {\bibfnamefont {L.}~\bibnamefont
  {Landau}}\ and\ \bibinfo {author} {\bibfnamefont {E.}~\bibnamefont
  {Lifshitz}},\ }\href@noop {} {\bibfield  {journal} {\bibinfo  {journal}
  {Pergamon Press, Oxford}\ } (\bibinfo {year} {1971})}\BibitemShut {NoStop}%
\bibitem [{\citenamefont {Labun}\ and\ \citenamefont
  {Rafelski}(2010)}]{labun2010dark}%
  \BibitemOpen
  \bibfield  {author} {\bibinfo {author} {\bibfnamefont {L.}~\bibnamefont
  {Labun}}\ and\ \bibinfo {author} {\bibfnamefont {J.}~\bibnamefont
  {Rafelski}},\ }\href@noop {} {\bibfield  {journal} {\bibinfo  {journal}
  {Physical Review D}\ }\textbf {\bibinfo {volume} {81}},\ \bibinfo {pages}
  {065026} (\bibinfo {year} {2010})}\BibitemShut {NoStop}%
\bibitem [{\citenamefont {Narozhny}\ and\ \citenamefont
  {Fofanov}(2000)}]{narozhny2000scattering}%
  \BibitemOpen
  \bibfield  {author} {\bibinfo {author} {\bibfnamefont {N.}~\bibnamefont
  {Narozhny}}\ and\ \bibinfo {author} {\bibfnamefont {M.}~\bibnamefont
  {Fofanov}},\ }\href@noop {} {\bibfield  {journal} {\bibinfo  {journal}
  {Journal of Experimental and Theoretical Physics}\ }\textbf {\bibinfo
  {volume} {90}},\ \bibinfo {pages} {753} (\bibinfo {year} {2000})}\BibitemShut
  {NoStop}%
\bibitem [{\citenamefont {Narozhny}\ \emph {et~al.}(2004)\citenamefont
  {Narozhny}, \citenamefont {Bulanov}, \citenamefont {Mur},\ and\ \citenamefont
  {Popov}}]{narozhny2004e+}%
  \BibitemOpen
  \bibfield  {author} {\bibinfo {author} {\bibfnamefont {N.}~\bibnamefont
  {Narozhny}}, \bibinfo {author} {\bibfnamefont {S.}~\bibnamefont {Bulanov}},
  \bibinfo {author} {\bibfnamefont {V.}~\bibnamefont {Mur}},\ and\ \bibinfo
  {author} {\bibfnamefont {V.}~\bibnamefont {Popov}},\ }\href@noop {}
  {\bibfield  {journal} {\bibinfo  {journal} {Physics Letters A}\ }\textbf
  {\bibinfo {volume} {330}},\ \bibinfo {pages} {1} (\bibinfo {year}
  {2004})}\BibitemShut {NoStop}%
\bibitem [{\citenamefont {Radier}\ \emph {et~al.}(2022)\citenamefont {Radier},
  \citenamefont {Chalus}, \citenamefont {Charbonneau}, \citenamefont
  {Thambirajah}, \citenamefont {Deschamps}, \citenamefont {David},
  \citenamefont {Barbe}, \citenamefont {Etter}, \citenamefont {Matras},
  \citenamefont {Ricaud} \emph {et~al.}}]{radier202210}%
  \BibitemOpen
  \bibfield  {author} {\bibinfo {author} {\bibfnamefont {C.}~\bibnamefont
  {Radier}}, \bibinfo {author} {\bibfnamefont {O.}~\bibnamefont {Chalus}},
  \bibinfo {author} {\bibfnamefont {M.}~\bibnamefont {Charbonneau}}, \bibinfo
  {author} {\bibfnamefont {S.}~\bibnamefont {Thambirajah}}, \bibinfo {author}
  {\bibfnamefont {G.}~\bibnamefont {Deschamps}}, \bibinfo {author}
  {\bibfnamefont {S.}~\bibnamefont {David}}, \bibinfo {author} {\bibfnamefont
  {J.}~\bibnamefont {Barbe}}, \bibinfo {author} {\bibfnamefont
  {E.}~\bibnamefont {Etter}}, \bibinfo {author} {\bibfnamefont
  {G.}~\bibnamefont {Matras}}, \bibinfo {author} {\bibfnamefont
  {S.}~\bibnamefont {Ricaud}}, \emph {et~al.},\ }\href@noop {} {\bibfield
  {journal} {\bibinfo  {journal} {High Power Laser Science and Engineering}\
  }\textbf {\bibinfo {volume} {10}},\ \bibinfo {pages} {e21} (\bibinfo {year}
  {2022})}\BibitemShut {NoStop}%
\bibitem [{\citenamefont {Eckey}\ \emph {et~al.}(2022)\citenamefont {Eckey},
  \citenamefont {Voitkiv},\ and\ \citenamefont {M{\"u}ller}}]{eckey2022strong}%
  \BibitemOpen
  \bibfield  {author} {\bibinfo {author} {\bibfnamefont {A.}~\bibnamefont
  {Eckey}}, \bibinfo {author} {\bibfnamefont {A.}~\bibnamefont {Voitkiv}},\
  and\ \bibinfo {author} {\bibfnamefont {C.}~\bibnamefont {M{\"u}ller}},\
  }\href@noop {} {\bibfield  {journal} {\bibinfo  {journal} {Physical Review
  A}\ }\textbf {\bibinfo {volume} {105}},\ \bibinfo {pages} {013105} (\bibinfo
  {year} {2022})}\BibitemShut {NoStop}%
\bibitem [{\citenamefont {Bulanov}\ \emph {et~al.}(2011)\citenamefont
  {Bulanov}, \citenamefont {Esirkepov}, \citenamefont {Hayashi}, \citenamefont
  {Kando}, \citenamefont {Kiriyama}, \citenamefont {Koga}, \citenamefont
  {Kondo}, \citenamefont {Kotaki}, \citenamefont {Pirozhkov}, \citenamefont
  {Bulanov} \emph {et~al.}}]{bulanov2011design}%
  \BibitemOpen
  \bibfield  {author} {\bibinfo {author} {\bibfnamefont {S.}~\bibnamefont
  {Bulanov}}, \bibinfo {author} {\bibfnamefont {T.~Z.}\ \bibnamefont
  {Esirkepov}}, \bibinfo {author} {\bibfnamefont {Y.}~\bibnamefont {Hayashi}},
  \bibinfo {author} {\bibfnamefont {M.}~\bibnamefont {Kando}}, \bibinfo
  {author} {\bibfnamefont {H.}~\bibnamefont {Kiriyama}}, \bibinfo {author}
  {\bibfnamefont {J.}~\bibnamefont {Koga}}, \bibinfo {author} {\bibfnamefont
  {K.}~\bibnamefont {Kondo}}, \bibinfo {author} {\bibfnamefont
  {H.}~\bibnamefont {Kotaki}}, \bibinfo {author} {\bibfnamefont
  {A.}~\bibnamefont {Pirozhkov}}, \bibinfo {author} {\bibfnamefont
  {S.}~\bibnamefont {Bulanov}}, \emph {et~al.},\ }\href@noop {} {\bibfield
  {journal} {\bibinfo  {journal} {Nuclear Instruments and Methods in Physics
  Research Section A: Accelerators, Spectrometers, Detectors and Associated
  Equipment}\ }\textbf {\bibinfo {volume} {660}},\ \bibinfo {pages} {31}
  (\bibinfo {year} {2011})}\BibitemShut {NoStop}%
\end{thebibliography}%

\end{document}